\documentclass[%
 aip,
 jmp,%
 amsmath,amssymb,
 reprint,%
]{revtex4-2}

\usepackage{hyperref}
\usepackage{graphicx}
\usepackage{dcolumn}
\usepackage{bm}

\begin{document}

\preprint{AIP/123-QED}

\title{Worldvolume origin of Higher M Theories}

\author{Pinak Banerjee}
\email{banerjee.pinak30@gmail.com}
\affiliation{Department of Theoretical Physics, Tata Institute of Fundamental Research, Homi Bhabha Rd,
Mumbai 400005, India}

\date{\today}

\begin{abstract}
Exceptional Periodicity (EP) has taught us that there are families of M Theory-like superalgebras in spacetime dimensions $11,19,27,\dots$ up to infinity. In this paper, we make the conjecture that M Theory at each level of EP can be realized as a brane worldvolume theory of an M Theory superalgebra at some higher level of EP.
\end{abstract}

\maketitle

\section{Introduction}
After Witten's proposal of M Theory \cite{Witten:1995ex} in $10+1$ dimensions as the unification of all five superstring theories, attempts to go beyond 11 dimensions have been put forth by Vafa's F Theory \cite{Vafa:1996xn} in $10+2$ and S Theory \cite{Bars:1996ab} by Bars in $11+2$ dimensions. Sezgin unified the superalgebras for various String/M theories by formulating Super Yang-Mills (SYM) in $11+3$ dimensions \cite{Rudychev:1997ui}. Nishino later extended these superalgebras in groups of four families up to an arbitrarily large number of even spacetime dimensions \cite{Nishino:1997hk}. Susskind came up with the idea of bosonic M theory in $26+1$ dimensions \cite{Horowitz:2000gn}. Rios, Chester, and Marrani used Exceptional Periodicity to explore Nishino's superalgebras but with maximal time dimensions restricted to $4$ \cite{Rios:2018lhc}. They also conjectured that $10+1$ dimensional M Theory can be realized as the worldvolume theory of the 10-brane in $26+1$ dimensional M Theory-like superalgebra, which can be embedded inside the superalgebra in $D=27+3$ \cite{Marrani:2020qmg} \cite{Rios:2019rfc}. In this paper, we write the M algebra and try to make the conjecture that this holds for any level of EP, i.e. M Theory at any level of EP can be realized on the brane worldvolume of a certain M Theory at some higher level of EP.   More detailed discussions regarding EP can be found in \cite{Truini:2017jiy}.

\section{ The Brane Worldvolume Realization of M-like Theories}
The four families of SYMs based on global minimal, chiral $(1,0)$ superalgebras exist in spacetime dimensions $(9+8\textbf{n},1), (10+8\textbf{n},2), (11+8\textbf{n},3), (12+8\textbf{n},4)$; the superalgebras are given explicitly in equations (3.5)-(3.8) of \cite{Rios:2018lhc}. The superalgebras in more than one time dimension are not standard, since they do not contain the momentum generator. For our purpose, we explicitly state the second type for minimal chiral $\mathcal{N}=1$ $D = (10+8\textbf{n},2)$:
\begin{eqnarray}
\left\{ Q_{\alpha },Q_{\beta }\right\} &=&\frac{1}{2}\left( \Gamma ^{{\mu%
}_{1}{\mu}_{2}}C^{-1}\right) _{\alpha \beta }Z_{\left[ {\mu}_{1}{%
\mu}_{2}\right] }^{(2)}+\frac{1}{6!}\left( \Gamma ^{{\mu}_{1}...{\mu}%
_{6}}C^{-1}\right) _{\alpha \beta }Z_{\left[ {\mu}_{1}...{\mu}_{6}%
\right] }^{(6)}  \notag \\
&&+...\frac{1}{(6+4\textbf{n})!}\left( \Gamma ^{{\mu}_{1}...{\mu}_{6+4\textbf{n}}}C^{-1}\right)
_{\alpha \beta }Z_{\left[ {\mu}_{1}...{\mu}_{6+4\textbf{n}}\right] }^{(6+4\textbf{n})}.
\label{central-28}
\end{eqnarray}%
Now, upon reducing this on a timelike circle, we write the $\mathcal{N}=1$ M algebra in dimensions $D=(10+8\textbf{n},1)$:
  (here $\alpha ,\beta =1,...,2^{5+4\textbf{n}}$, whereas $\mu $-indices run
$0,1,...,10+8\textbf{n}$)
\begin{eqnarray}
\left\{ Q_{\alpha },Q_{\beta }\right\} &=&\left( \Gamma ^{\mu }C^{-1}\right)
_{\alpha \beta }P_{\mu }+\frac{1}{2}\left( \Gamma ^{\mu _{1}\mu
_{2}}C^{-1}\right) _{\alpha \beta }\underset{\text{M2}}{Z_{\left[ \mu
_{1}\mu _{2}\right] }^{(2)}}+ \dots \\ &&+\frac{1}{(5+4\textbf{n})!}\left( \Gamma ^{\mu _{1}...\mu
_{5+4\textbf{n}}}C^{-1}\right) _{\alpha \beta }\underset{\text{M(5+4\textbf{n})}}{Z_{\left[ \mu
_{1}...\mu _{5+4\textbf{n}}\right] }^{(5+4\textbf{n})}}  \notag \\
\end{eqnarray}
In $D=(10+8\textbf{n},1)$, we know ${Z_{\left[ \mu
_{1}...\mu _{5+4\textbf{n}}\right] }^{(5+4\textbf{n})}}\sim {Z_{\left[ \mu
_{1}...\mu _{6+4\textbf{n}}\right] }^{(6+4\textbf{n})}}$, so M$(6+4\textbf{n})$ brane is not present in the M algebra.

\medskip

 Now, we start with $10+1$ dimensional M Theory and want to see its worldvolume realization.
  The superalgebra for $D=10+1$ $\mathcal{N}=1$ is
  (here $\alpha ,\beta =1,...,32$, whereas $\mu $-indices run
$0,1,...,10$)
\begin{eqnarray}
\left\{ Q_{\alpha },Q_{\beta }\right\} &=&\left( \Gamma ^{\mu }C^{-1}\right)
_{\alpha \beta }P_{\mu }+\frac{1}{2}\left( \Gamma ^{\mu _{1}\mu
_{2}}C^{-1}\right) _{\alpha \beta }\underset{\text{M2}}{Z_{\left[ \mu
_{1}\mu _{2}\right] }^{(2)}}+\frac{1}{5!}\left( \Gamma ^{\mu _{1}...\mu
_{5}}C^{-1}\right) _{\alpha \beta }\underset{\text{M5}}{Z_{\left[ \mu
_{1}...\mu _{5}\right] }^{(5)}}  \notag 
\end{eqnarray}
 The superalgebra for $D=26+1$ $\mathcal{N}=1$ is
  (here $\alpha ,\beta =1,...,8,192$, whereas $\mu $-indices run
$0,1,...,26$)
\begin{eqnarray}
\left\{ Q_{\alpha },Q_{\beta }\right\} &=&\left( \Gamma ^{\mu }C^{-1}\right)
_{\alpha \beta }P_{\mu }+\frac{1}{2}\left( \Gamma ^{\mu _{1}\mu
_{2}}C^{-1}\right) _{\alpha \beta }\underset{\text{M2}}{Z_{\left[ \mu
_{1}\mu _{2}\right] }^{(2)}}+\frac{1}{5!}\left( \Gamma ^{\mu _{1}...\mu
_{5}}C^{-1}\right) _{\alpha \beta }\underset{\text{M5}}{Z_{\left[ \mu
_{1}...\mu _{5}\right] }^{(5)}}  \notag \\
&&+\frac{1}{6!}\left( \Gamma ^{\mu _{1}...\mu _{6}}C^{-1}\right) _{\alpha
\beta }\underset{\text{M6}}{Z_{\left[ \mu _{1}...\mu _{6}\right] }^{(6)}}+%
\frac{1}{9!}\left( \Gamma ^{\mu _{1}...\mu _{9}}C^{-1}\right) _{\alpha \beta
}\underset{\text{M9}}{Z_{\left[ \mu _{1}...\mu _{9}\right] }^{(9)}}  \notag
\\
&&+\frac{1}{10!}\left( \Gamma ^{\mu _{1}...\mu _{10}}C^{-1}\right) _{\alpha
\beta }\underset{\text{M10}}{Z_{\left[ \mu _{1}...\mu _{10}\right] }^{(10)}}+%
\frac{1}{13!}\left( \Gamma ^{\mu _{1}...\mu _{13}}C^{-1}\right) _{\alpha \beta
}\underset{\text{M13}}{Z_{\left[ \mu _{1}...\mu _{13}\right] }^{(13)}}  \notag  
\end{eqnarray}

\medskip

 The central extensions in $D=26+1$ can be obtained by a Kaluza-Klein timelike-reduction from the $%
(1,0)$ minimal chiral superalgebra in $D=26+2$, whose central extensions
read with $\mathbf{n}=2$ ( $\hat{\mu}$-indices
here run $\mathring{0}$,$0,1,...,26$)%
\begin{eqnarray}
\left\{ Q_{\alpha },Q_{\beta }\right\} &=&\frac{1}{2}\left( \Gamma ^{\hat{\mu%
}_{1}\hat{\mu}_{2}}C^{-1}\right) _{\alpha \beta }Z_{\left[ \hat{\mu}_{1}\hat{%
\mu}_{2}\right] }^{(2)}+\frac{1}{6!}\left( \Gamma ^{\hat{\mu}_{1}...\hat{\mu}%
_{6}}C^{-1}\right) _{\alpha \beta }Z_{\left[ \hat{\mu}_{1}...\hat{\mu}_{6}%
\right] }^{(6)}  \notag \\
&&+\frac{1}{10!}\left( \Gamma ^{\hat{\mu}_{1}...\hat{\mu}_{10}}C^{-1}\right)
_{\alpha \beta }Z_{\left[ \hat{\mu}_{1}...\hat{\mu}_{10}\right] }^{(10)}+%
\frac{1}{14!}\left( \Gamma ^{\hat{\mu}_{1}...\hat{\mu}_{14}}C^{-1}\right)
_{\alpha \beta }Z_{\left[ \hat{\mu}_{1}...\hat{\mu}_{14}\right] }^{(14)}.
\end{eqnarray}%
By splitting $\hat{\mu}=\mathring{0},\mu $, one indeed obtains%
\begin{eqnarray}
Z_{\hat{\mu}_{1}\hat{\mu}_{2}}^{(2)} &\rightarrow &\left\{
\begin{array}{l}
Z_{\mu _{1}\mathring{0}}^{(2)}\sim P_{\mu }; \\
Z_{\mu _{1}\mu _{2}}^{(2)};%
\end{array}%
\right. \\
Z_{\hat{\mu}_{1}...\hat{\mu}_{6}}^{(6)} &\rightarrow &\left\{
\begin{array}{l}
Z_{\mu _{1}...\mu _{5}\mathring{0}}^{(5)}\sim Z_{\mu _{1}...\mu _{5}}^{(5)};
\\
Z_{\mu _{1}...\mu _{6}}^{(6)};%
\end{array}%
\right. \\
Z_{\hat{\mu}_{1}...\hat{\mu}_{10}}^{(10)} &\rightarrow &\left\{
\begin{array}{l}
Z_{\mu _{1}...\mu _{9}\mathring{0}}^{(10)}\sim Z_{\mu _{1}...\mu _{9}}^{(9)};
\\
Z_{\mu _{1}...\mu _{10}}^{(10)},%
\end{array}%
\right. \\
Z_{\hat{\mu}_{1}...\hat{\mu}_{14}}^{(14)} &\rightarrow &\left\{
\begin{array}{l}
Z_{\mu _{1}...\mu _{13}\mathring{0}}^{(14)}\sim Z_{\mu _{1}...\mu
_{13}}^{(13)}; \\
Z_{\mu _{1}...\mu _{14}}^{(14)}\rightarrow \epsilon _{\mu _{1}...\mu
_{27}}Z_{\nu _{14}...\nu _{27}}^{(14)}\eta ^{\mu _{14}\nu _{14}}...\eta
^{\mu _{27}\nu _{27}}\sim Z_{\mu _{1}...\mu _{13}}^{(13)},%
\end{array}%
\right.
\end{eqnarray}%
        Therefore one obtains $D=10+1$ M Theory as worldvolume theory of the M10-brane in $D=26+1$ superalgebra as discussed in \cite{Marrani:2020qmg}.

        \medskip

        Now we ask the following question. Can this $D=26+1$ theory be realized as the worldvolume theory of some brane? 

We turn our attention to the superalgebra for $\mathcal{N}=1$ $D=  50+2$ whose central extensions read $\mathbf{n}=5$ ( $\hat{\mu}$-indices
here run $\mathring{0}$,$0,1,...,50$)%
\begin{eqnarray}
\left\{ Q_{\alpha },Q_{\beta }\right\} &=&\frac{1}{2}\left( \Gamma ^{\hat{\mu%
}_{1}\hat{\mu}_{2}}C^{-1}\right) _{\alpha \beta }Z_{\left[ \hat{\mu}_{1}\hat{%
\mu}_{2}\right] }^{(2)}+\frac{1}{6!}\left( \Gamma ^{\hat{\mu}_{1}...\hat{\mu}%
_{6}}C^{-1}\right) _{\alpha \beta }Z_{\left[ \hat{\mu}_{1}...\hat{\mu}_{6}%
\right] }^{(6)}  \notag \\
&&+....+\frac{1}{26!}\left( \Gamma ^{\hat{\mu}_{1}...\hat{\mu}_{26}}C^{-1}\right)
_{\alpha \beta }Z_{\left[ \hat{\mu}_{1}...\hat{\mu}_{26}\right] }^{(26)}.
\end{eqnarray}%
By reduction on a timelike circle, we get for $D= 50+1$ that $Z_{\mu_1..\mu_{26}}^{(26)}\sim Z_{\mu_1..\mu_{25}}^{(25)}$, and thus we cannot get the $26+1$ dimensional worldvolume brane; we get only the M25 brane in $D=50+1$. 

\medskip

Now, we move onto the $\mathcal{N}=1$, $D=58+2$ superalgebra with  $\mathbf{n}=6$ ( $\hat{\mu}$-indices here run $\mathring{0}$,$0,1,...,58$)%
\begin{eqnarray}
\left\{ Q_{\alpha },Q_{\beta }\right\} &=&\frac{1}{2}\left( \Gamma ^{\hat{\mu%
}_{1}\hat{\mu}_{2}}C^{-1}\right) _{\alpha \beta }Z_{\left[ \hat{\mu}_{1}\hat{%
\mu}_{2}\right] }^{(2)}+\frac{1}{6!}\left( \Gamma ^{\hat{\mu}_{1}...\hat{\mu}%
_{6}}C^{-1}\right) _{\alpha \beta }Z_{\left[ \hat{\mu}_{1}...\hat{\mu}_{6}%
\right] }^{(6)}  \notag \\
&&+....+\frac{1}{26!}\left( \Gamma ^{\hat{\mu}_{1}...\hat{\mu}_{26}}C^{-1}\right)
_{\alpha \beta }Z_{\left[ \hat{\mu}_{1}...\hat{\mu}_{26}\right] }^{(26)}++\frac{1}{30!}\left( \Gamma ^{\hat{\mu}_{1}...\hat{\mu}_{30}}C^{-1}\right)
_{\alpha \beta }Z_{\left[ \hat{\mu}_{1}...\hat{\mu}_{30}\right] }^{(30)}.
\end{eqnarray}%
The superalgebra for $D=58+1$ $\mathcal{N}=1$ is
  (here $\alpha ,\beta =1,...,2^{29}$, whereas $\mu $-indices run
$0,1,...,58$)
\begin{eqnarray}
\left\{ Q_{\alpha },Q_{\beta }\right\} &=&\left( \Gamma ^{\mu }C^{-1}\right)
_{\alpha \beta }P_{\mu }+\frac{1}{2}\left( \Gamma ^{\mu _{1}\mu
_{2}}C^{-1}\right) _{\alpha \beta }\underset{\text{M2}}{Z_{\left[ \mu
_{1}\mu _{2}\right] }^{(2)}}+\frac{1}{5!}\left( \Gamma ^{\mu _{1}...\mu
_{5}}C^{-1}\right) _{\alpha \beta }\underset{\text{M5}}{Z_{\left[ \mu
_{1}...\mu _{5}\right] }^{(5)}}  \notag \\
&&+\frac{1}{6!}\left( \Gamma ^{\mu _{1}...\mu _{6}}C^{-1}\right) _{\alpha
\beta }\underset{\text{M6}}{Z_{\left[ \mu _{1}...\mu _{6}\right] }^{(6)}}+...+%
\frac{1}{25!}\left( \Gamma ^{\mu _{1}...\mu _{25}}C^{-1}\right) _{\alpha \beta
}\underset{\text{M25}}{Z_{\left[ \mu _{1}...\mu _{25}\right] }^{(25)}}  \notag 
\\
&&+\frac{1}{26!}\left( \Gamma ^{\mu _{1}...\mu _{26}}C^{-1}\right) _{\alpha
\beta }\underset{\text{M26}}{Z_{\left[ \mu _{1}...\mu _{26}\right] }^{(26)}}+%
\frac{1}{29!}\left( \Gamma ^{\mu _{1}...\mu _{29}}C^{-1}\right) _{\alpha \beta
}\underset{\text{M29}}{Z_{\left[ \mu _{1}...\mu _{29}\right] }^{(29)}}  \notag 
\\.  
\end{eqnarray}
Upon reduction on a timelike circle to $D=58+1$, $Z_{\left[ \hat{\mu}_{1}...\hat{\mu}_{26}\right] }^{(26)}$ breaks up into $Z_{\left[ \hat{\mu}_{1}...\hat{\mu}_{26}\right] }^{(26)}$ and $Z_{\left[ \hat{\mu}_{1}...\hat{\mu}_{25}\right] }^{(25)}$; this M26 brane with worldvolume signature $26+1$ can now be used as realization of the $D=26+1$ dimensional M Theory.
The $D=26+1$ M Theory contains P, M2, M5, M6, M9, M10, and M13 branes (as can be seen by dimensional reduction of 2, 6, 10, 14-branes in $D=26+2$); while the $D=58+1$ M Theory contains P, M2, M5, M6, M9, M10, M13, M14, M17, M18, M21, M22, M25, M26, M29 branes (as can be seen by dimensional reduction of 2, 6, 10, 14, 18, 22, 26, 30-branes in $D=58+2$). Thus $D=26+1$ M Theory can be realized as the M26 Brane worldvolume theory. 

\medskip

This $D=58+1$ M Theory can now be realized as worldvolume theory on M58 Brane arising in the superalgebra of $\mathcal{N}=1$, $D= 122+1$; which again can be realized as worldvolume theory on M122 Brane present in  $\mathcal{N}=1$, $D= 250+1$; which in turn finds its origin on the worldvolume of M250 Brane in $\mathcal{N}=1$, $D= 506+1$ and so on. (The earlier levels naively seem to contain the required branes, but they are ruled out due to the duality relations relating $Z_{\left[ \hat{\mu}_{1}...\hat{\mu}_{p}\right] }^{(p)}$ and $Z_{\left[ \hat{\mu}_{1}...\hat{\mu}_{D-p}\right] }^{(D-p)}$ in D dimensions that we encountered for $D= 50+1$ too). The full spectrum of branes can be obtained easily.

\medskip

This chain has captured only a few levels of EP. What about the $D=18+1$ M Theory?

The superalgebra for $D=18+1$ $\mathcal{N}=1$ is
  (here $\alpha ,\beta =1,...,512$, whereas $\mu $-indices run
$0,1,...,18$)
\begin{eqnarray}
\left\{ Q_{\alpha },Q_{\beta }\right\} &=&\left( \Gamma ^{\mu }C^{-1}\right)
_{\alpha \beta }P_{\mu }+\frac{1}{2}\left( \Gamma ^{\mu _{1}\mu
_{2}}C^{-1}\right) _{\alpha \beta }\underset{\text{M2}}{Z_{\left[ \mu
_{1}\mu _{2}\right] }^{(2)}}+\frac{1}{5!}\left( \Gamma ^{\mu _{1}...\mu
_{5}}C^{-1}\right) _{\alpha \beta }\underset{\text{M5}}{Z_{\left[ \mu
_{1}...\mu _{5}\right] }^{(5)}}  \notag \\
&&+\frac{1}{6!}\left( \Gamma ^{\mu _{1}...\mu _{6}}C^{-1}\right) _{\alpha
\beta }\underset{\text{M6}}{Z_{\left[ \mu _{1}...\mu _{6}\right] }^{(6)}}+%
\frac{1}{9!}\left( \Gamma ^{\mu _{1}...\mu _{9}}C^{-1}\right) _{\alpha \beta
}\underset{\text{M9}}{Z_{\left[ \mu _{1}...\mu _{9}\right] }^{(9)}}  \notag
\end{eqnarray}

\medskip

We turn to $\mathcal{N}=1$ $D=42+2$ superalgebra with with  $\mathbf{n}=4$ ( $\hat{\mu}$-indices here run $\mathring{0}$,$0,1,...,42$)%
\begin{eqnarray}
\left\{ Q_{\alpha },Q_{\beta }\right\} &=&\frac{1}{2}\left( \Gamma ^{\hat{\mu%
}_{1}\hat{\mu}_{2}}C^{-1}\right) _{\alpha \beta }Z_{\left[ \hat{\mu}_{1}\hat{%
\mu}_{2}\right] }^{(2)}+\frac{1}{6!}\left( \Gamma ^{\hat{\mu}_{1}...\hat{\mu}%
_{6}}C^{-1}\right) _{\alpha \beta }Z_{\left[ \hat{\mu}_{1}...\hat{\mu}_{6}%
\right] }^{(6)}  \notag \\
&&+....+\frac{1}{18!}\left( \Gamma ^{\hat{\mu}_{1}...\hat{\mu}_{18}}C^{-1}\right)
_{\alpha \beta }Z_{\left[ \hat{\mu}_{1}...\hat{\mu}_{18}\right] }^{(18)} \\ &&+\frac{1}{22!}\left( \Gamma ^{\hat{\mu}_{1}...\hat{\mu}_{22}}C^{-1}\right)
_{\alpha \beta }Z_{\left[ \hat{\mu}_{1}...\hat{\mu}_{22}\right] }^{(22)}.
\end{eqnarray}%
The superalgebra for $D=42+1$ $\mathcal{N}=1$ is
  (here $\alpha ,\beta =1,...,2^{21}$, whereas $\mu $-indices run
$0,1,...,42$)
\begin{eqnarray}
\left\{ Q_{\alpha },Q_{\beta }\right\} &=&\left( \Gamma ^{\mu }C^{-1}\right)
_{\alpha \beta }P_{\mu }+\frac{1}{2}\left( \Gamma ^{\mu _{1}\mu
_{2}}C^{-1}\right) _{\alpha \beta }\underset{\text{M2}}{Z_{\left[ \mu
_{1}\mu _{2}\right] }^{(2)}}+\frac{1}{5!}\left( \Gamma ^{\mu _{1}...\mu
_{5}}C^{-1}\right) _{\alpha \beta }\underset{\text{M5}}{Z_{\left[ \mu
_{1}...\mu _{5}\right] }^{(5)}}  \notag \\
&&+\frac{1}{6!}\left( \Gamma ^{\mu _{1}...\mu _{6}}C^{-1}\right) _{\alpha
\beta }\underset{\text{M6}}{Z_{\left[ \mu _{1}...\mu _{6}\right] }^{(6)}}+...+%
\frac{1}{17!}\left( \Gamma ^{\mu _{1}...\mu _{17}}C^{-1}\right) _{\alpha \beta
}\underset{\text{M17}}{Z_{\left[ \mu _{1}...\mu _{17}\right] }^{(17)}}  \notag 
\\
&&+\frac{1}{18!}\left( \Gamma ^{\mu _{1}...\mu _{18}}C^{-1}\right) _{\alpha
\beta }\underset{\text{M18}}{Z_{\left[ \mu _{1}...\mu _{18}\right] }^{(18)}}+%
\frac{1}{21!}\left( \Gamma ^{\mu _{1}...\mu _{21}}C^{-1}\right) _{\alpha \beta
}\underset{\text{M21}}{Z_{\left[ \mu _{1}...\mu _{21}\right] }^{(21)}}  \notag .  
\end{eqnarray}
Upon timelike reduction to $\mathcal{N}=1$ $D=42+1$ we get $Z_{\left[ \hat{\mu}_{1}...\hat{\mu}_{18}\right] }^{(18)}$ spilitting into $Z_{\left[ \hat{\mu}_{1}...\hat{\mu}_{17}\right] }^{(17)}$ and $Z_{\left[ \hat{\mu}_{1}...\hat{\mu}_{18}\right] }^{(18)}$.
The $D=18+1$ M Theory contains P, M2, M5, M6, and M9 branes (as can be seen by dimensional reduction of 2, 6, 10-branes in $D=18+2$); while the $D=42+1$ M Theory contains P, M2, M5, M6, M9, M10, M13, M14, M17, M18, and M21 branes (as can be seen by dimensional reduction of 2, 6, 10, 14, 18, 22-branes in $D=42+2$). Thus $D=42+1$ M Theory contains M18 brane corresponding to $Z_{\left[ \hat{\mu}_{1}...\hat{\mu}_{18}\right] }^{(18)}$. This M18 brane with worldvolume signature $18+1$ can be used as a realization of $\mathcal{N}=1$ $D=18+1$ M Theory. Again, the $\mathcal{N}=1$ $D=42+1$ theory can be realized as the worldvolume of M42 brane in $\mathcal{N}=1$ $D=90+1$ superalgebra, and this goes on similarly.

\medskip

Motivated by the earlier examples, we want to see this realization on a generic level. Say, $10+8\textbf{n}= 26+16\textbf{n'}$; then $\textbf{n} = 2+2\textbf{n'}$. Now, looking at the superalgebra in $D=(10+8\textbf{n},2)$, we see the branes present are $2, 6, ..., 2+4\textbf{n}, 6+4\textbf{n}$ corresponding to the central charges which translated to  \textbf{n'} language read $2, 6, ..., 10+8\textbf{n'}, 14+8\textbf{n'}$ branes. Upon reduction on a timelike circle, we get $Z_{\left[ \hat{\mu}_{1}...\hat{\mu}_{10+8\textbf{n'}}\right] }^{(10+8\textbf{n'})}$ splitting into $Z_{\left[ \hat{\mu}_{1}...\hat{\mu}_{10+8\textbf{n'}}\right] }^{(10+8\textbf{n'})}$ and $Z_{\left[ \hat{\mu}_{1}...\hat{\mu}_{9+8\textbf{n'}}\right] }^{(9+8\textbf{n'})}$. The  $Z_{\left[ \hat{\mu}_{1}...\hat{\mu}_{10+8\textbf{n'}}\right] }^{(10+8\textbf{n'})}$ corresponds to a M(10+8\textbf{n'}) brane in $D=(26+16\textbf{n'})+1$ which can serve as a worldvolume realization of $D=(10+8\textbf{n'})+1$ M Theory. 

\medskip

Thus, we come to the generic worldvolume realization:

\emph{M Theory in $D=(10+8\textbf{n},1)$ can be realized as a worldvolume theory of M~$(10+8\textbf{n})$ brane present in $\mathcal{N}=1$ superalgebra of $D=(26+16\textbf{n},1)$. Here $\textbf{n} = 0,1,2,\dots,\infty$}

\section{Finer discussions}
\begin{itemize}
    \item We evade Nahm's no-go susy beyond $10+1$ dimensions (assuming single time dimension) in the following way: we have to get down from $26+1$ to $10+1$ dimensions by some non-trivial brane compactification as discussed in \cite{Marrani:2020qmg}, and then from $10+1$, we go to $3+1$ dimensions by standard Kaluza Klein reduction. Upon this reduction and suitable truncation, the lagrangian (5.17)-(5.25) of \cite{Marrani:2020qmg} reduces to the lagrangian of $D=10+1$ $\mathcal{N}=1$ supergravity (sugra).
    \item Whether these minimal chiral $\mathcal{N}=1$ superalgebras admit physical theories with local Susy remains to be seen; though a candidate lagrangian for $\mathcal{N}=1$ sugra in $D= 26+1$ has been outlined in \cite{Marrani:2020qmg}; it is formidable to prove that it is supersymmetric! The massless spectrum of the would-be sugra theories might be obtained from the corresponding ESYM theories by double copy.
    \item From the braneworld perspective, p-brane worldvolumes give rise to spacetimes, they induce spacetime signatures.
    \end{itemize}
    \section{Conclusion}
    In this work, we have conjectured that all the higher M theories can be realized as some brane worldvolume theory, though the existence of actual local susy in these higher M theories remains to be seen. Also, we would like to extend these EP results and try to explore worldvolume realization to Hull's M*, M' theory originally developed in $9+2$ and $6+5$ dimensions \cite{Hull:1998ym}.


\bibliography{ref}

\end{document}